\documentstyle[12pt]{article}
\begin{document}
%New Commands
\newcommand{\siml}{\stackrel{<}{\sim}}
\newcommand{\simg}{\stackrel{>}{\sim}}
\baselineskip=1.2\baselineskip

%single vs. double space 
\noindent
\begin{center}
{\large\bf
Spike-Train Responses of a Pair of Hodgkin-Huxley Neurons \\
with Time-Delayed Couplings 
} 
\end{center}

\begin{center}
Hideo Hasegawa
%\footnote{e-mail:  hasegawa@u-gakugei.ac.jp}
\end{center}

\begin{center}
{\it Department of Physics, Tokyo Gakugei University  \\
Koganei, Tokyo 184-8501, Japan}
\end{center}
\begin{center}
{\rm (November 10, 2000)}
\end{center}
%\date{\today}
%\maketitle
\thispagestyle{myheadings}

\begin{center} 
{\bf Abstract}   \par
\end{center} 
Model calculations have been performed on the spike-train 
response of a pair of Hodgkin-Huxley (HH) neurons coupled 
by recurrent excitatory-excitatory couplings  
with time delay. 
The coupled, excitable HH neurons are assumed to receive
the two kinds of spike-train inputs: 
the transient input consisting of
$M$ impulses for the finite duration ($M$: integer)
and the sequential
input with the constant interspike interval (ISI).
The distribution of the 
output ISI $T_{\rm o}$ shows a rich of variety
depending on the coupling strength and the time delay.
The comparison is made between
the dependence of the output ISI for the transient
inputs and that for the sequential inputs.

\vspace{1.5cm}

\vspace{1.5cm}

\noindent
{\it Corresponding Author}

\noindent  
Hideo Hasegawa \\ 
Department of Physics, Tokyo Gakugei University \\ 
4-1-1 Nukui-kita machi, Koganei, Tokyo 184-8501, Japan \\
Phone: 042-329-7482, Fax: 042-329-7491 \\
e-mail: hasegawa@u-gakugei.ac.jp

%\baselineskip=1.5\baselineskip

%
\newpage
\section{Introduction}

Neurons communicate by producing sequence of action 
potentials or spikes. It has been widely believed 
that information is encoded in the average rate of firings, 
the number of action potentials over some suitable 
intervals.
This firing rate hypothesis was first proposed by 
Andrian \cite{Andrian26} from a study of frog, in which
the firing rate monotonically increases with an increase
of the stimulus strength.
By applying the firing rate hypothesis, the properties
of many types of neurons in brain have been investigated
and the theoretical models have been 
developed \cite{Hopfield82}.

When all action potentials are taken to be identical
and only the times of firing of a given neuron
are considered, we obtain a discrete series of times,
$\{t_{n}\}$, which is expected to contain the information.
In the {\it rate coding}, only the average of the rate of
the interspike interval (ISI) is taken into account, and
then some or most of this information is neglected.

In recent years, the alternative {\it temporal coding},
in which detailed spike timing is taken to play an important role,
is supported by experiments in a variety of 
biological systems: 
sonar processing of bats \cite{Suga83}, 
sound localization of owls \cite{Konishi92},
electrosensation in electric fish \cite{Carr86}, 
visual processing
of cats \cite{Eckhorn88}\cite{Gray89}, 
monkeys \cite{Rolls94} and human \cite{Thorpe96}.
It is now primarily important to understand what
kind of code is employed in biological systems:
rate code, temporal code or others
\cite{Rieke96}\cite{Maass98}.

Neural functions are performed in the activity of neurons.
Since the Hodgkin-Huxley (HH) model was proposed to account for
the squid giant axon \cite{Hodgkin52}, 
its property has been intensively
investigated. 
Its responses to applied dc 
\cite{Nemoto76}-\cite{Guttman80}
and sinusoidal currents \cite{Aihara84}\cite{Matsumoto84} 
have been studied.
The HH-type models have been widely employed for a study on 
activities of {\it transducer neurons} such as motor
and relay neurons, 
which transform
amplitude-modulated inputs to spike-train outputs.
Regarding the single HH neuron as a {\it data-processing neuron},
the present author \cite{Hasegawa00} 
(referred to as I hereafter) has investigated its response
to the spike-train inputs whose ISIs
are modulated by deterministic, semi-deterministic (chaotic)
and stochastic signals.

Several investigations have been reported on
the property of a pair of the HH neurons
\cite{Hansel93}-\cite{Kanamaru98}.
In the network of two HH oscillators coupled 
by excitatory couplings without time delay, 
the unit fires periodically in the
synchronized state.
%just as in the fully connected
%large network \cite{Strogatz91}.
It is, however, not the case when the excitatory
couplings have some time delay,
for which the anti-phase state becomes more stable
than the synchronized state \cite{Vreeswijk94}.
Rather, inhibitory couplings with substantial time delay
lead to the in-phase synchronized states
in the coupled HH oscillators \cite{Vreeswijk94}.
The similar conclusion is obtained also in the coupled
integrate-and-fire (IF) oscillators [22,29,31-34]. 
%\cite{Vreeswijk94, Ernst98}
%\cite{Ernst95}-\cite{Coobes97}
The phase diagrams for the synchronized state
and various cluster states in the coupled HH oscillators
are obtained as functions of the synapse
strength and the time delay [23-26]. 
%\cite{Park96}-\cite{Kim98}
%\cite{Park96}-\cite{Yoshinaga99}.
Recurrent loops involving two or more neurons with 
excitatory and/or inhibitory synapses 
are found in biological systems
such as hippocampus \cite{Freud96}, 
neo-cortex \cite{Kisvarday93} and 
thalamus \cite{Bal93}.
It is important to make a detailed study on the
coupled HH neurons, which is the simplest but 
meaningful network unit.

In recent years, much attention has been paid
to the delayed-feedback systems described by the 
delay-differential equation (DDE) [38-43].
%\cite{Ikeda82}-\cite{Fisher94}
Their property has been investigated with the use of
various functional forms for the delay-feedback term in DDE. 
The exposed properties include 
the odd-harmonic solutions \cite{Ikeda82}\cite{Ikeda87},
the bifurcation leading to chaos [39-41],
%\cite{Ikeda87, Mork90} 
the multistability \cite{Ikeda87},
and the chaotic itinerancy \cite{Otsuka90}\cite{Fisher94}.
Among them the multistability is intrigue because it
may be one of conceivable mechanisms for memory storage
in biological neural networks.
It has been shown by Ikeda and Matsumoto\cite{Ikeda87} 
that when the delay time is larger
than the response time in the delayed feedback system,
information may be stored in temporal patterns. 
Actually, Foss, Longtin, Mensour and Milton \cite{Foss96}
demonstrate this ability in the coupled HH (and IF) neurons
with the time-delayed feedback.

The response of DDE is usually discussed by applying the
sequential sinusoidal or spike-train inputs
to non-linear systems.
In real neural systems, however, it is not so often for 
neurons to receive
such sequential, continuous inputs. Rather, it is 
expected to be more realistic that neurons receive
clustered inputs including information to be processed.
The purpose of the present paper is to investigate
the response of the coupled, excitable
HH neurons to both the transient and sequential
spike-train inputs.
We adopt the recurrent excitatory-excitatory (E-E)
couplings between a pair of HH neurons,
to which we apply
the transient spike-train inputs consisting
of clustered $M$ impulses for the finite duration ($M$: integer)
as well as the sequential inputs with the
constant ISI.

Our paper is organized as follows:
In the next $\S 2$, we describe a simple neural system
consisting of neurons, axons, synapses and dendrites, which is 
adopted for our numerical calculation.
We present the calculated results in $\S 3$:
the response of the coupled HH neurons to
the transient, clustered impulses is discussed in $\S 3.1$
and that to the sequential spike-train input in $\S 3.2$. 
The dependence of the 
distribution of the output ISIs on the coupling strength
and the time delay are studied.
The final $\S 4$ is devoted to conclusion and discussion.

\section{Adopted Model}

We adopt a simple neural system
consisting of a pair of neurons which is numbered 1 and 2.
The neurons which are described by the HH model with
identical parameters,
are coupled with the time delay
of $\tau_{jk} \: (j,k=1,2)$ for an impulse 
propagating from the neuron $k$ 
to the neuron $j$.
This delay time is the sum of conduction times
through the axon and dendrite.
It has been reported that real biological synapses
exhibit temporal dynamics of depression or
potentiation during neuronal computation
\cite{Abbott97}\cite{Tsodyks98}.
We, however, treat the synapse as a static
unit for a simplification of our calculation.
The synapse with the coupling strength $C_{jk}$ 
is excitatory, and
it is assumed to be described by the alpha function [eq. (7)].

Dynamics of the membrane potential $V_{j}$ of
the coupled HH neuron {\it j} (=1, 2) 
is described by the non-linear DDEs given by 
\begin{eqnarray}
\bar{C} {\rm d} V_{j}(t)/{\rm d} t &=& -I_{j}^{\rm ion}
(V_{j}, m_{j}, h_{j}, n_{j})  
+ I_{j}^{\rm ext} \nonumber \\
&& + I_{j}^{\rm int}(\{V_{k}(t-\tau_{jk})\}),
\end{eqnarray}
%\begin{equation}
%\bar{C} d V_{j}(t)/d t = -I_{j}^{\rm ion}(V_{j}, m_{j}, h_{j}, n_{j})  
%+ I_{j}^{\rm ext} + I_{j}^{\rm int}(\{V_{k}(t-\tau_{jk})\}),
%\end{equation}
%
where $\bar{C} = 1 \; \mu {\rm F/cm}^2$ is the capacity of the membrane.
The first term of eq. (1) expresses the ion current given by
\begin{eqnarray}
&& I_{j}^{\rm ion}(V_{j}, m_{j}, h_{j}, n_{j}) 
= g_{\rm Na} \:m_{j}^3 \:h_{j} \:(V_{j} - V_{\rm Na}) \nonumber \\
&& + g_{\rm K} \:n_{j}^4 \:(V_{j} - V_{\rm K}) 
+ g_{\rm L} \:(V_{j} - V_{\rm L}).
\end{eqnarray}
%\begin{equation}
%I_{j}^{\rm ion}(V_{j}, m_{j}, h_{j}, n_{j}) 
%= g_{\rm Na} m_{j}^3 h_{j} (V_{j} - V_{\rm Na})
%+ g_{\rm K} n_{j}^4 (V_{j} - V_{\rm K}) 
%+ g_{\rm L} (V_{j} - V_{\rm L}).
%\end{equation}
Here the maximum values of conductivities 
of Na and K channels and leakage are
$g_{\rm Na} = 120 \; {\rm mS/cm}^2$, 
$g_{\rm K} = 36 \; {\rm mS/cm}^2$ and
$g_{\rm L} = 0.3 \; {\rm mS/cm}^2$, respectively; 
the respective reversal potentials are   
$V_{\rm Na} = 50$ mV, $V_{\rm K} = -77$ mV and 
$V_{\rm L} = -54.5 $ mV.
The gating variables of Na and
K channels, $m_{j}, h_{j}$ and $n_{j}$,
are described by 
\begin{equation}
{\rm d} m_{j}/{\rm d} t = - (a_{mj} + b_{mj}) \: m_{j} + a_{mj},
\end{equation}
\begin{equation}
{\rm d} h_{j}/{\rm d} t = - (a_{hj} + b_{hj}) \: h_{j} + a_{hj},
\end{equation}
\begin{equation}
{\rm d} n_{j}/{\rm d} t = - (a_{nj} + b_{nj}) \: n_{j} + a_{nj}.
\end{equation}
The coefficients of 
$a_{mj}$ and $b_{mj}$ {\it etc.} are expressed in terms of 
$V_{j}$ (their explicit expressions 
having been given in refs. \cite{Hasegawa00}\cite{Park96}) and
then the variables $V_{j}$, $m_{j}$, $h_{j}$ 
and $n_{j}$ are coupled. 

The second term in eq. (1) denotes the external input currents given by
\begin{equation}
I_{j}^{\rm ext} = I_{{\rm s}j} + A_{s} \:\delta_{j1} \sum_{n} 
\alpha(t-t_{{\rm i}n}),
\end{equation}
with the alpha function $\alpha(t)$ given by
\begin{equation}
\alpha(t) = (t/\tau_{\rm s}) \; {\rm e}^{-t/\tau_{\rm s}} \:  \Theta(t).
\end{equation}
The first term ($I_{{\rm s}j}$) in eq. (6) is the dc current which determines
whether the neuron is excitable or periodically oscillating.
Its second term expresses the postsynaptic current
which is induced by
the  presynaptic spike-train input applied 
to the neuron 1, given by
\begin{equation}
U_{\rm i}(t) = V_{\rm a} \: \sum_n  \: \delta (t - t_{{\rm i}n}).
\end{equation}
In eqs. (2.6)-(2.8),
$\Theta (t)=1$ for $x \geq 0$ and 0 for $x < 0$;
$A_{\rm s}=g_{\rm s} \:(V_{\rm a}-V_{\rm s})$,
$g_{\rm s}$ and $V_{\rm s}$ stand 
for the conductance and reversal potential,
respectively, of the synapse;
$\tau_{\rm s}$ is the time constant relevant 
to the synapse conduction, which is assumed to be
$\tau_{\rm s}=2$ msec; 
$t_{{\rm i}n}$ is the $n$-th firing time of 
the spike-train inputs defined recurrently by
\begin{equation}
t_{{\rm i}n+1} = t_{{\rm i}n} + T_{{\rm i}n}(t_{{\rm i}n}),
\end{equation}
where the input ISI $T_{{\rm i}n}$ is generally a function
of $t_{{\rm i}n}$.
For the constant input ISI of $T_{{\rm i}n}=T_{\rm i}$, 
$t_{in}$ is given by
$t_{{\rm i}n} = n \:T_{\rm i}$ 
for an integer $n$.

When the membrane potential of the {\it j}-th neuron $V_{j}(t)$ oscillates,
it yields the spike-train output, which may be expressed by
\begin{equation}
U_{{\rm o}j}(t) = V_{\rm a} \: \sum_m \: \delta (t - t_{{\rm o}jm}),
\end{equation}
in a similar form to eq. (8), 
$t_{{\rm o}jm}$ being the $m$-th firing time of the neuron $j$
when $V_{j}(t)$ crosses $V_{z}$ = 0.0 mV from below.
The output ISI is given by
\begin{equation}
T_{{\rm o}jm} = t_{{\rm o}jm+1} - t_{{\rm o}jm}.
\end{equation}

The third term in eq. (1) which 
expresses the interaction between the two neurons,
is assumed to be given by
\begin{equation}
I_{j}^{\rm int}(\{V_{k}(t-\tau_{jk})\})
=   \sum_{k (\neq j)} \sum_{m} \: C_{jk} 
\: \alpha(t-\tau_{jk}-t_{{\rm o}km}).
%=  A_{s} \sum_{k (\neq j)} \sum_{m} \: c_{jk} 
%\: \alpha(t-\tilde{t}_{{\rm o}km}).
\end{equation}
%Note that $\tilde{t}_{{\rm o}km}$ (=$t_{{\rm o}km}+\tau_{jk}$) is
%a function of  $V_{j}(t-\tau_{jk})$ 
%defined by $V_{j}(\tilde{t}_{{\rm o}jm}-\tau_{jk}-0) < V_{z}$ 
%and $V_{j}(\tilde{t}_{{\rm o}jm}-\tau_{jk}+0) \geq V_{z}$.
We assume the recurrent excitatory-excitatory couplings with
positive $C_{ij}$ given by 
$\mid C_{21} \mid =\mid C_{12} \mid \equiv c \: A_{\rm s}$ and 
$\tau_{21}=\tau_{12} \equiv \tau_{\rm d}$.

As for the functional form of the coupling term of
$I_{j}^{\rm int}(\{V_{k}(t-\tau_{jk})\})$,
Foss, Longtin, Mensour and Milton \cite{Foss96}
adopt a simpler form given by 
\begin{equation}
I_{j}^{\rm int}(\{V_{k}(t-\tau_{jk})\})
=   \sum_{k (\neq j)}  \: \mu_{jk} \: V_{k}(t-\tau_{jk}),
\end{equation}
taking no account of the synapse, 
where $\mu_{jk}$ is the coefficient of the synaptic coupling.
They discuss the memory storage of the pattern in output spike trains,
injecting the input information by the initial function,
$V(t)$ for $t \in [-\tau_{\rm d}, 0)$, whereas in our calculation
input information is given by $I^{\rm ext}_{j}$ [eq. (6)].

Differential equations given by eqs. (1)-(5) including
the external current and 
couplings given by eqs. (6)-(12) are solved 
by the forth-order Runge-Kutta method.
%and partly
%by the exponential method \cite{Wilson98}
%for a confirmation of the accuracy.
The calculation for each set of parameters is performed
for 2 sec (200,000 steps) 
with the integration time step of 0.01 msec
with double precision.
The initial conditions for the variables are given by
\begin{eqnarray}
&& V_{j}(t)= -65 \:\: \mbox{\rm mV}, m_{j}(t)=0.0526,  
h_{j}(t)=0.600, \nonumber \\
&& n_{j}(t)=0.313, \:\: \mbox{\rm for} \:\: j=1, 2
\:\:  \mbox{at} \: t=0,
\end{eqnarray}
which are the rest-state solution of a single
HH neuron ($C_{jk}=0$).
The initial function for $V_{j}(t)$,
whose setting is indispensable for
the delay-differential equation, is given by
\begin{equation}
V_{j}(t)= -65 \:\: \mbox{\rm mV} 
\:\: \mbox{\rm for} \:\: j=1,2
\:\:  \mbox{at} \: t \in [-\tau_{\rm d}, 0).
\end{equation}
For an analysis of asymptotic solutions,
we discard results of initial 1000 msec
(100,000 steps).

\section{Calculated Resuls}

In the present study, we consider only the excitable HH neurons
by setting $I_{{\rm s}j}=0$ and 
$A_{\rm s}=g_s \:(V_a-V_s)=40 \mu {\rm A/cm}^2$
for $g_s=0.5 {\rm mS/cm}^2$, $V_a=30$ and $V_s=-50$ mV \cite{Hasegawa00}.
Our model has additional three parameters, $T_{\rm i}$, 
$\tau_{\rm d}$ and $c$. We treat them as free parameters to be changed 
because the values of ISI and the time delay observed
in biological systems distribute 
in a fairly wide range \cite{note1}.

\subsection{Transient Spike-Train Inputs }

Let us first investigate the response to the transient,
clustered spike-train
inputs consisting of $M$ impulses.
We have studied in I,  
the transient response of a single HH neuron
to spike-train inputs consisting of $M=2-5$ impulses 
with $T_{\rm i}=5, 10$ and 20 msec (see  Fig. 20 of I).  
In the case of $T_{\rm i}=20$ msec, we get $T_{\rm o}=20$ msec 
and  the number of output
pulses is the same as that of input pulses.  On the contrary,
in the cases of $T_{\rm i}=5$ and 10 msec, the ISI of output is generally
larger than that of input because of its character of the low-pass filter,
and the number of output pulse is not necessary the
same as that of input pulse.

Figure 1 shows the example of
the time courses of input ($U_{\rm i}$),
output pulses ($U_{{\rm o}j}$),
the total postsynaptic current 
($I_{j}=I_{j}^{ext}+I_{j}^{int}$) and the membrane potential ($V_{j}$)
with $M=3$, $T_{\rm i}=20$, $\tau_{\rm d}=10$ msec and $c=1.0$
for the E-E coupling ($c > 0$).
The first external pulse
applied at $t=0$ yields the firing of the neuron 1
after the intrinsic delay of $\tau_{{\rm i}1} \sim 2$ msec. 
The emitted impulse propagates the axon and 
reaches the synapse of the neuron 2 after $\tau_{21}=10$ msec.
After a more delay of an intrinsic  $\tau_{{\rm i}2} \sim 2$ msec,
the neuron 2 makes the firing which 
yields the input current to the neuron 1 
after a delay of $\tau_{12}=10$ msec.
The input pulses trigger the continuous oscillation in
the coupled HH neurons with the output
ISI of $T_{{\rm o}}=24.10$ msec. The time dependence of
the output ISI of the neuron 1 and 2 are plotted
by solid and dashed lines in Fig. 2(a), respectively.
We note that $T_{{\rm o}1}$ and  $T_{{\rm o}2}$ start from 
the values of 20.00 and 19.96 msec, respectively, 
and soon become the value of 24.10 msec.

Figures 2(b) and 2(c) show similar plots for different
values of $\tau_{\rm d}$= 13.75 and 20 msec, respectively.
In the case of $\tau_{\rm d}$=13.75 msec, output ISIs start
the oscillation with $T_{\rm o}$=20.00 and 12.36 msec and
asymptotically approach the value of 15.93 msec.
On the contrary, in the case of $\tau_{\rm d}$=20 msec,
the oscillation of $T_{\rm o}$ starting at $t=0$ continues
with the asymptotic values of $T_{\rm o}$= 18.51 and 25.59 msec.
In the following subsections,
we will discuss the dependence of output ISIs on the time delay and the
coupling strength.
Since its behavior of the spike-train outputs of 
the neurons 1 and 2 is similar, 
we hereafter take into account only that of the neuron 1 
otherwise noticed.

\vspace{0.5cm}

\noindent
%\begin{center}
\subsubsection{The time-delay dependence}
%\end{center}

Now we study how the output ISIs are determined.
When the coupling strength is sufficiently strong
for inputs to trigger output impulses and when the
feedback time $T_{\rm fb}$ is larger than the duration
of clustered impulses
({\it i.e.} $T_{\rm fb}=2 \tau_{\rm d} + \tau_{{\rm i}1} + \tau_{{\rm i}2} 
> (M-1) \:T_{\rm i}  $), we get two values of
$T_{{\rm o}}$ given by
\begin{eqnarray}
&& T^{(1)}_{\rm o} = T_{\rm i}, \nonumber \\
&& T^{(2)}_{\rm o} = T_{\rm fb} - (M-1) \:T_{\rm i} \nonumber \\
&& = 2 \tau_{\rm d} + \tau_{{\rm i}1} + \tau_{{\rm i}2} 
- (M-1) \:T_{\rm i}.    
\end{eqnarray}
On the other hand, when the feedback time is shorter than
input-pulse duration ($2 \tau_{\rm d} + \tau_{{\rm i}1} + \tau_{{\rm i}2} 
< (M-1) \:T_{\rm i}$),
we get
\begin{eqnarray}
&& T^{(1)}_{\rm o} = T_{\rm i} \;\Theta(M-3), \nonumber \\ 
&& T^{(2)}_{\rm o} = \mid \ell \:T_{\rm fb} - m \:T_{\rm i} \mid \nonumber \\
&& = \mid \ell\: (2 \tau_{\rm d} + \tau_{{\rm i}1} + \tau_{{\rm i}2}) 
- m \:T_{\rm i} \mid,  
\end{eqnarray}
where integers $\ell$ and $m$ satisfy 
$1 \leq \ell \leq [(M-1)T_{\rm i}/T_{\rm fb}]+1$ and
$0 \leq m \leq M-1$, $[\:\cdot\:]$ is the Gauss sign 
and $T^{(1)}_{\rm o}$ is vanishing for $M \leq 2$.
%In the above consideration,
%we do not take into account the effects of the absolute
%refractory period and the merging of external and recurrent 
%inputs,
%which make its calculation of output ISI more complicated 
%as will be shown below.

Figures 3(a) and 3(b) show
the calculated time-delay dependence of $T_{\rm o}$ for $c=1.0$
and 1.6, respectively.
Filled and open circles denote $T_{\rm o}$s of
the transient ($t \leq 1000$ msec) and 
asymptotic solutions ($t > 1000$ msec), respectively.
As was shown in Figs. 2(a)-2(c),
the output ISIs of the asymptotic solutions
are $T_{\rm o}$=24.10 for $\tau_{\rm d}$= 10 msec, $T_{\rm o}$=15.93  
for $\tau_{\rm d}$=13.75 msec, and $T_{\rm o}$=18.51 and 25.59 msec 
for $\tau_{\rm d}$=20 msec. 
We note that the behavior of $T_{\rm o}$ strongly
depends on the value of $\tau_{\rm d}$.
In order to see their detailed structures, we show
in Figs. 4(a)-4(c), enlarged plots of the
narrow regions in Fig. 3(b).
Figures 3(a) and 3(b) show three main branches expressed by
$T_{\rm o} \sim 2 \tau_{\rm d} +5$, $T_{\rm o} \sim 2 \tau_{\rm d}-15$
and $T_{\rm o} \sim 2 \tau_{\rm d}-35$, which are
obtainable for a pair of integers of $(\ell, m)$
=(1,0), (1,1) and (1,2), respectively,
by eq. (17) with $\tau_{{\rm i}1}=\tau_{{\rm i}2}=2.5$
and $T_{\rm i}=20$ msec. 
Figure 4(a), in which the narrow region of
$10 \leq \tau_{\rm d} \leq 20$ msec in Fig. 3(b) is enlarged,
shows an additional  branch of a single ISI given by 
$T_{\rm o} \sim \tau_{\rm d} +2$ at $11.9 < \tau_{\rm d} < 12.3$ 
and $13.7 < \tau_{\rm d} \siml 19$ msec.
Furthermore we note in Fig. 4(b) which shows an enlarged
plot at $20 \leq \tau_{\rm d} \leq 30$ msec in Fig. 3(b), 
a branch of multiple ISIs given by
$T_{\rm o} \sim 0.5\tau_{\rm d} +5$ at $21.9 < \tau_{\rm d} < 28.5$ msec. 
These $\tau_{\rm d}$ dependences of $T_{\rm o}$ 
cannot be explained by eq. (17), and may be harmonics
of the fundamental ISI with the period of $2\tau_{\rm d}$.
The $\tau_{\rm d}$ dependence of $T_{\rm o}$ for $c=1.6$
shown in Fig. 3(b) is similar to that for $c=1.0$
shown in Fig. 3(a), except an additional branch
given by $T_{\rm o} \sim 0.5 \tau_{\rm d} +4$ 
at $12.3 < \tau_{\rm d} < 13.7$ msec.
The narrow region of $16  \leq \tau_{\rm d} \leq 20$ msec in Fig. 4(a) 
is enlarged in Fig. 4(c), where the ISI of 
the asymptotic solution shows 
the stair-like structure.

\vspace{0.5cm}

\noindent
%\begin{center}
\subsubsection{The coupling-strength dependence}
%\end{center}

Figures 5(a) and 5(b) show the $c$ dependence of the
output ISI of the transient (filled circles)
and asymptotic solutions (open circles).
As was shown in Figs. 2(a) and 2(b),
the ISI of the asymptotic solutions with $c=1.0$
is $T_{\rm o}$=24 msec for $\tau_{\rm d}=10$ msec and 
$T_{\rm o}$=15.93 msec for $\tau_{\rm d}$=13.75 msec. 
Figure 5(a) shows that as the coupling strength becomes weak, 
$T_{\rm o}$ is increased because of the integrator character
of the HH neuron.
A similar effect is obtained also in a single HH
neuron, in which the output ISI becomes larger for 
smaller spike-train inputs \cite{Hasegawa00}.
Figure 5(b) shows that
ISIs for the transient solutions fluctuate 
around that for the asymptotic solution as expected.
The enlarged plot for $1.5 \leq c \leq 1.7$ of Fig. 5(b)
is given in Fig. 6, where a discontinuous change in $T_{\rm o}$ 
is clearly realized at $c=1.61$ msec.
For $c < 0.2$ neurons emit only three impulses,
returning to rest without oscillations.

\subsection{ Sequential Spike-Train Inputs}

Next we discuss the response to the sequential spike-train.
Our calculations in I show that
when an isolated HH neuron ($c=0$) receives the sequential
inputs with the constant ISI of $T_{\rm i}$,
it behaves as a low-pass filter:
it emits the spike train with $T_{\rm o}>10$ msec
for $T_{\rm i}<12$ msec while for $T_{\rm i}>12$ msec
its output ISI is given by $T_{\rm o}=T_{\rm i}$
(see Fig. 7 of ref. 20).
This response may be modified when the coupling
is introduced to a pair of HH neurons.
Figure 7 shows
the time courses of input ($U_{\rm i}$),
output ($U_{{\rm o}j}$),
the total postsynaptic current 
($I_{j}=I_{j}^{\rm ext}+I_{j}^{\rm int}$) and the 
membrane potential ($V_{j}$)
for $T_{\rm i}=20$ msec, $\tau_{\rm d}=10$ msec and $c=1.0$,
which are the same  parameters adopted for the
clustered inputs shown in Fig. 1.
The output ISI in Fig. 1 is 24.1 msec while that
in Fig. 7 is 20 msec which is the entrained value
with input ISI.
The response behavior of the coupled neurons strongly 
depends on the parameters of $c$, $\tau_{\rm d}$ and $T_{\rm i}$.

\vspace{0.5cm}

\noindent
%\begin{center}
\subsubsection{The time-delay dependence}
%\end{center}

Figures 8(a) and 8(b) show the $\tau_{\rm d}$ dependence of the
distribution of $T_{\rm o}$ for $c=1.0$ and 1.6, respectively,
in the asymptotic solution
of the sequential inputs \cite{note2}.
The calculations in Figs. 8(a) and 8(b) are performed with the same
parameters of $T_{\rm i}$ and $c$ in Figs. 3(a) and 3(b), 
respectively. 
The $\tau_{\rm d}$ dependence of $T_{\rm o}$ in Fig. 8(a) [Fig. 8(b)]
is quite different from that in Fig. 3(a) [Fig. 3(b)].
Figures 8(a) and 8(b) have the bifurcation structure,
as commonly observed in systems with the delayed feedback
\cite{Ikeda87}.
In order to see more the detailed structure
of the bifurcation,
we show, in Fig. 9, the enlarged plot  
for the range of $21 \leq \tau_{\rm d} \leq 26$ msec
between the dotted, vertical lines in Fig. 8(a).
The region sandwiched by verical dotted lines in Fig. 9(a)
($21 \leq \tau_{\rm d} \leq 26$ msec)
is further enlarged in Fig. 9(b).
Figures 9(a) and 9(b) clearly show 
the bifurcation as changing $\tau_{\rm d}$.
The $\tau_{\rm d}$ dependence for $c=1.6$ shown in Fig. 8(b)
is similar to that for
$c=1.0$ shown in Fig. 8(a),
and its enlarged plot also exhibits the bifurcation (not shown).

\vspace{0.5cm}

\noindent
%\begin{center}
\subsubsection{The coupling-strength dependence}
%\end{center}

The calculated $c$ dependence of the
distribution of $T_{\rm o}$ with
$\tau_{\rm d}=10$ and 13.75 msec for $T_{\rm i}=20$ msec
are shown in Figs. 10(a) and 10(b), respectively \cite{note2}.
The adopted values of $T_{\rm i}$ and $\tau_{\rm d}$ 
in  Figs. 10(a) and 10(b) are the same as
those in Figs. 5(a) and 5(b), respectively.
The output ISI for the sequential inputs shown
in Fig. 10(a) is 20 msec ($= T_{\rm i}$)
independent of the coupling constant,
while that for the clustered inputs shown in Fig. 5(a)
decreases monotonically as the $c$ value is decreased.
We note in Fig. 10(b) that, as increasing the $c$ value,
the distribution of the output ISIs for $\tau=13.75$ msec
exhibits the bifurcation.
In order to investigate the phenomenon in more detail,
we show, in Fig. 11, the enlarged plot 
for the range of $0.6 \leq c \leq 1.2$
sandwiched by the dotted, vertical lines in Fig. 10(b).

A cycle whose output ISIs almost
continuously distribute, is expected to be chaotic 
although in the strict sense,
the distribution of our $T_{\rm o}$s never 
becomes continuous because they
are {\it quantized} by the integration time step of 0.01 msec.
Among many candidates of chaos-like behavior in Figs. 11, 
we pay our attention
to the result of $c=0.95$, for which
the Lorentz plot (return map) of its $T_{\rm o}$ is shown 
in Fig. 12(a)
(calculations are performed for 20 sec of two million steps).
The output ISIs seem to distribute on the folded ring.
When these points are connected by lines in the temporal
order, the inside of the ring is nearly filled by them.
In order to examine the property of this cycle,
we calculate the correlation dimension $\nu$ given by
\cite{Grassberger83}
\begin{equation}
\nu= \lim_{\epsilon \rightarrow 0} 
\frac{\log \: C(\epsilon)}{\log \: \epsilon},
\end{equation}
with
\begin{equation}
C(\epsilon)=N^{-2} \sum_{m,n=1}^{N} 
\Theta(\epsilon - \mid \mbox{\boldmath $X_{m}$}
- \mbox{\boldmath $X_{n}$}   \mid),
\end{equation}
\begin{equation}
\mbox{\boldmath $X_{m}$}= (T_{{\rm o}m}, \: T_{{\rm o}m+1}, 
\:...., \: T_{{\rm o}m+k-1} ),
\end{equation}
where $C(\epsilon)$ is the correlation integral, {\boldmath $X_{m}$}
is the $k$-dimensional vector generated by $T_{{\rm o}m}$,
$N$ the size of data, and $\Theta(\cdot)$ the Heaviside function.
Figure 12(b) shows the $\log C(\epsilon)$-$\log \epsilon$ plot
for various embedding dimensions $k$ 
calculated for the cycle shown in Fig. 12(a)
with the data size of $N \sim 1200$.
%where $k$ is the embedding
%dimension, $C(\epsilon)$ the correlation integral
%and $\epsilon$ the distance.
We note that $C(\epsilon)$ behaves as 
$C(\epsilon) \propto \epsilon^{\nu}$ with the correlation
dimension of $\nu =0.94\pm0.02$
for small $\epsilon \:\: (0.01 = e^{-4.6} < \epsilon < e^{0})$.
When the relevant spike-train output given by $U_{{\rm o}1}(t)$ [eq. (10)] 
is Fourier transformed, it spectrum shows a broad distribution. 
%The non-integral $\nu$
%and a broad Fourier spectrum
These suggest that the cycle shown in Fig. 12(a) may be chaos, 
although we cannot draw any definite conclusion 
until a detailed calculation of its Lyapunov spectrum is performed, 
related discussion being given in $\S 4$.

\vspace{0.5cm}
\section{Conclusion and Discussion}

We have performed model calculations of the
spike-train responses of a pair of coupled HH neurons,
applying the two types of inputs of
the transient and  sequential spike-train impulses.
Calculations for the transient inputs shown in  
Figs. 3(a) and 3(b) [Fig. 5(a) and 5(b)]
are performed with the same model parameters as those for
the sequential inputs shown in 
Figs. 8(a) and 8(b) [10(a) and 10(b)].
When we make a comparison of the response to transient 
inputs with the corresponding result to sequential inputs,
we notice the difference and similarity between them.
When we regard a neuron as a data-processing element,
the relation between input and output ISIs is one of
the important factors for its quality.
Our previous calculation in I shows that a single HH neuron emits
a single ISI of $T_{\rm o}=T_{\rm i}$ for $T_{\rm i} < 12$ msec 
whereas for shorter ISI of $T_{\rm i} < 12$ msec it emits multiple 
ISIs of $T_{\rm o} > 10$ msec (see Fig. 7 of I).
Figures 13(a) and 13(b) show the $T_{\rm i}$ dependence of
$T_{\rm o}$ of coupled HH neurons
for the transient and sequential inputs, respectively,
with $\tau_{\rm d}=50$ msec and $c=1.0$.
Dashed lines in Figs. 13(a) and 13(b) are obtained
with the use of eq. (17) for a pair of 
integers $(\ell, m)$ shown in the brackets. 
It is apparent that
the distribution of $T_{\rm o}$ in Fig. 13(a) is not
the same as that in Fig. 13(b), but they are partly similar.
%For example, we obtain for $T_{\rm i}=20$ msec, 
%$T_{\rm o}=19.6$, 20.0 and 64.54 msec
%in Fig. 13(a) whereas only 20.0 msec in Fig. 13(b).

On the theoretical point of view, the sequential input
is taken as the limit of $M \rightarrow \infty$
of the $M$-impulse clustered input.
In order to understand the transition of the response 
behavior as increasing $M$, we plot, in Fig. 14,
the time dependence of $T_{\rm o}$ for this set
of parameters by changing the $M$ value.
For $M=3$, $T_{\rm o}$ oscillates with the values of
19.6, 20.0 and 64.54 msec, 
as mentioned before.
The calculated $T_{\rm o}$ for $M$=4 are 18.7, 19.8, 20.0, and 45.6 msec, 
and those for $M=5$ are 19.9, 20.0 and 24.2 msec.
For $M=10$, $T_{\rm o}$ remains 20 msec until $t \sim 200$ msec, after which
$T_{\rm o}$ oscillates with the values of
19.9, 20.0 and 24.2 msec.
In the limit of $M \rightarrow \infty$ corresponding to the 
sequential inputs, the state with $T_{\rm o}=20$ msec
continues from $t=0$ to $\infty$.
Thus as increasing $M$, the time region of $T_{\rm o}=20$ msec
is increased.
Figure 15 shows the similar plot of the time
dependence of the distribution of $T_{\rm o}$
for various $M$
with $T_{\rm i}=20$, $\tau_{\rm d}=13.75$ and $c=0.95$,
for which the sequential input leads to the chaotic behavior,
as was discussed in Sec. 3.2 (see Fig. 12(a)).
In the case of $M=3$, we get the oscillation in $T_{\rm o}$
which asymptotically approaches the value of 15.97 msec.
In the case of $M=10$ (50), the chaotic behavior is
realized at $0 \leq t \siml 180$
($0 \leq t \siml 980$) msec during the application of inputs.
After inputs are switched off,
the output ISI gradually approaches the asymptotic value of
15.97 msec.
In the limit of $M \rightarrow \infty$, the chaotic oscillation
eternally continues.

We have shown in $\S 3.2$ that 
the cycle of the output ISIs shown
in Fig. 12(a) may be chaos because of
its correlation dimension of $\nu \sim 0.94$
derived from the $\log C(\epsilon)$-$\log \epsilon$ 
plot in Fig. 12(b).
This is not surprising because the response of
{\it single} HH neurons to some kinds of
external inputs may be chaotic
\cite{Aihara84}\cite{Matsumoto84}\cite{Hasegawa00}.
%\cite{Aihara84, Matsumoto84, Hasegawa00}
In particular, it has been shown in I that the response of 
a single HH neuron may be chaos when  the ISI of
the spike-train input is modulated by
the sinusoidal signal: \cite{Hasegawa00}
\begin{equation}
T_{\rm i}(t) = d_0 + d_1 \: \sin(2 \pi t/T_p),
\end{equation}
where $d_0$ denotes the average of $T_{\rm i}(t)$, 
$d_1$ the magnitude of the sinusoidal modulation, 
and $T_{p}$ its period.
Figure 16(a) shows the Lorentz plot of the output 
ISIs of the single HH neuron ($c=0.0$) receiving 
sequential inputs modulated by sinusoidal ISIs
[eq. (21)] with $d_0=2 d_1=20$ and $T_p=100$ msec 
(see Fig. 9(d) of I, where points in the Lorentz plot are
connected by lines in the temporal order).
We note that $T_{\rm o}$s distribute on the deformed ring.
From the $\log C(\epsilon)$-$\log \epsilon$ plot (not shown)
of this cycle, we get its correlation dimension
of $\nu \sim 1.04$.
We apply this sinusoidal spike-train input to the coupled
HH neurons with $\tau_{\rm d}=10$ msec and $c=1.0$, 
whose Lorentz plot is shown in Fig. 16(b).
Its structure is rather different from that shown 
in Fig. 16(a). 
Actually the correlation dimension of this cycle for
the coupled HH neurons
is $\nu \sim 1.83$, which is different from and larger than
$\nu \sim 1.04$ of the cycle shown in Fig. 16(a) 
for the single HH neuron.  
From similar calculations for the coupled HH neurons, 
we obtain the correlation dimensions of
$\nu \sim 0.95$ for $\tau_{\rm d}=5$ msec and $c=1.0$, and
$\nu \sim 1.03$ for $\tau_{\rm d}=10$ msec and $c=0.5$.
These results clearly show that
the correlation dimension of the output ISIs
depend not only 
on the model parameters ($c$ and $\tau_{\rm d}$)
of the coupled HH neurons but also 
on $\nu_{\rm i}$, the correlation dimension
of input ISIs ($\nu_{\rm i}=0$ for the constant ISI and
$\nu_{\rm i}=1$ for the sinusoidally modulated ISI).
We expect that spike-train inputs with larger $\nu_{\rm i}$
lead to spike-train outputs with larger $\nu$.
One of the disadvantages of the present calculation
of the correlation dimension
is a lack of the data size of $N \sim 1200$
with million-step calculations.
A more accurate analysis  
requires a larger size of data and then
a computer with the larger 
memory storage.

Next we discuss the time correlation 
$\Gamma_{12}(\tau)$ between the 
membrane potentials, $V_1$ and $V_2$, of the 
neurons 1 and 2, defined by

\begin{equation}
\Gamma_{12}(\tau) = \int_{t_a}^{t_b} 
[V_1(t)-<V_1(t)>] \: [V_2(t+\tau)-<V_2(t)>] \: {\rm d}t,
\end{equation}
where the bracket denotes the time average, and $t_a=1000$ 
and $t_b=2000$ msec are adopted for our calculation.
Figure 17(a) shows the result
for the case of the sequential input
to the coupled HH neurons with $T_{\rm i}=20$, 
$\tau_{\rm d}=10$ msec and $c=1.0$
(see Fig. 7 for the time courses of $V_1$ and $V_2$).
In this case we obtain the constant $T_{\rm o}=20$ msec as 
was discussed in $\S$ 3.2, and then
$\Gamma_{12}(\tau)$ shown in Fig. 17(a)
has peaks at $\tau=12.04+20 \:n$ msec ($n$: integer)
with the period of 20 msec, as expected.
We are interested in the time correlation
for the case when the distribution
of $T_{\rm o}$ is chaotic.
Results for such cases are shown in Figs. 17(b) and 17(c).
We have discussed in $\S$ 3.2 that the cycle of $T_{\rm o}$
depicted in Fig. 12 may be chaotic.
Figure 17(b) shows the result of this case
for the E-E coupling with $T_{\rm i}=20$, 
$\tau_{\rm d}=13.75$ msec and $c=0.95$.
We note that $\Gamma_{12}(\tau)$ has peaks
%at $\tau$=0.0, 16.07, $\sim 3.2$, 48.17, 62.71,...
at $\tau$=-45.34, -30.69, -16.13, 0.0, 16.07, 30.54, 48.17,...
msec with the period of about 16 msec, which is
the sum of $\tau_{\rm d}$ and $\tau_{i1}$.
More evident peaks are found in Fig. 17(c) showing also
the chaotic case discussed in the preceding paragraph: 
the E-E coupled
neurons receiving the sinusoidal inputs given by eq. (21) with 
$d_0=2 d_1=20$, $T_{\rm p}=100$, $T_{\rm i}=20$, 
$\tau_{\rm d}=10$ msec and $c=1.0$ [see Fig. 16(b)].
We note peaks in $\Gamma_{12}(\tau)$ 
%at $\tau$=0.0, 12.72, 25.30, 37.88, 50.43, ... msec 
at $\tau$=-37.63, -25.03, -12.52 0.0, 12.72, 25.30, 37.88, ... msec 
with the period of about 12.6 msec.
These results are not modified even when the initial condition of
one of the HH neurons is slightly changed from the values in eq. (14).
It is interesting that the synchronization is well
preserved between the coupled HH neurons 
in the chaotic state \cite{Hansel96}.

A fairly large variability ($c_{\rm v}=0.5 \sim 1.0$) has
been reported for spike trains of non-bursting cortical
neurons in V1 and MT of monkey \cite{Softky92}.
It is possible that when the appreciable 
variability in neuronal signals
is taken into account in our calculations,
much of the fine structures in the $c-$ and 
$\tau_{\rm d}$-dependent distributions of $T_{o}$ will be washed out.
In order to study this speculation, we apply the spike-train input
with ISI whose distribution is given by the gamma distribution
defined by \cite{Hasegawa00}
\begin{equation}
P(T) = s^r \;\; T^{r - 1} \;\; e^{- s T}/\; \Gamma(r),
\end{equation}
where $\Gamma \;(r)$ is the gamma function.
The average of input ISI is given by
$\mu_{\rm i} =  r/s$, its root-mean-square (RMS)  by
$\sigma_{\rm i} = \sqrt{ r}/s$ and its variability by
$c_{\rm vi} = \sigma_{\rm i}/\mu_{\rm i}=1/\sqrt{r}$.
Figure 18 shows the $\tau_{\rm d}$ dependence of 
the mean ($\mu_{\rm o}$) and 
RMS values ($\sigma_{\rm o}$) of the
output ISIs for $c_{\rm vi}=0.0$ (dashed curves)
and $c_{\rm vi}=0.43$ (solid curves) 
with $T_{\rm i}=20$ msec and $c=1.0$.
Note that $\sigma_{\rm o}$ provides us with the measure
of the width of the distribution of $T_{\rm o}$.
The distribution for $c_{\rm vi}=0$ has a fine
structure reflecting the strong $\tau_{\rm d}$ 
dependence of $T_{\rm o}$ [see Fig. 8(a)]. 
This fine structure is, however, 
washed out for $c_{\rm vi}=0.43$, as expected.

Finally we discuss the relevance of the calculated 
properties to biological experiments.
Many experimental data have shown
the complex behavior of 
electoencephalographic (EEG) waves in brain.
The macroscopic characteristics of their activity are
aperiodic and unpredicable oscillations with amplitude 
histograms that are near Gaussian, auto-correlation functions 
that rapidly approach zero and intermittent burst of oscillations
having spectral peaks \cite{Lopetz95}.
It has been reported that the activity of EEG in the olfactory bulb
shows the significance of chaos in an animal's motivated
behaviors \cite{Skarda87}.
The complex behavior of EEG is nothing but the reflection of that
of action potentials generated by neurons.
It has been shown that neurons in different regions of the brain
have different firings property.
In hippocampus, for example,
gamma oscillation ($20 \sim 70$ Hz) occurs in vivo, 
following sharp waves \cite{Freud96}.
In neo-cortex, gamma oscillation is observed under conditions of
sensory signal as well as during sleep \cite{Kisvarday93}.
In thalamus
%which is a gateway interposed between 
%the peripheral sensory areas and the neo-cortex, 
burst firings are found during slow-sleep, and single spiking 
is found during arousal \cite{Bal93}. 
One of the reasons of this variety of firings is
that different class of neurons has different ion conductances.
%Nevertheless, different responses patterns are realized
%for the same type of neurons in the same network.
Physiological experiments have shown 
that these biological systems include recurrent loops 
connecting two or more neurons
with excitatory and/or inhibitory synapses.
It is conceivable that the distributed processing of brain 
function may be due to differences not only in ion conductances
of the neuron
but also in synaptic strength and in delay times of axons and 
dendrites connecting neurons.
Although many theoretical studies have been made,
the origin of the complexity in neuron firings
has not been well clarified at the moment. 
We should note that synaptic strengths may be modified by Hebb's
learning rule, which changes the state of the network including 
given synapses.
Our calculations for a pair of HH neurons, which is
a simplest, plausible model simulating recurrently connected network,
have demonstrated that depending on the 
coupling strength and the time delay, the coupled HH neurons
show a much variety, yielding not only regular spike trains 
but also irregular (chaotic) impulses.
We hope that our calculations might have some relevance to 
the complex activities in real, biological systems.

\section*{Acknowledgements}
This work is partly supported by
a Grant-in-Aid for Scientific Research from the Japanese 
Ministry of Education, Science and Culture.

\newpage
%REFERENCES
%

%

\newpage

\vspace{0.5cm}

\noindent{\large\bf  Figure Captions}   \par

\vspace{0.5cm}

\noindent
{\bf Fig.1} 
The time dependence of 
the clustered input ($U_{\rm i}$), 
output ($U_{{\rm o}j}$),
the total postsynaptic current ($I_{j}$) and the
membrane potential ($V_{j}$) with $M=3$,
$T_{\rm i}=20$, $\tau_{\rm d}=10$ msec and $c=1.0$.
The result of $V_2$ is shifted
downward by 200 mV and scales for $U_{\rm i}$, 
$U_{{\rm o}j}$ and $I_j$ are arbitrary.

\vspace{0.5cm}

\noindent
{\bf Fig.2}
The time dependence of the output ISI ($T_{{\rm o}j}$)
of neuron 1 (filled circles) and 2 (open circles)
for (a) $\tau_{\rm d}=10$, (b) 13.75 and  (c) 20 msec
for clustered inputs
with $M=3$, $T_{\rm i}=20$ msec and $c=1.0$.

\vspace{0.5cm}

\noindent
{\bf Fig.3}
The $\tau_{\rm d}$ dependence of the
distribution of $T_{{\rm o}}$  
of (a) $c=1.0$ and (b) 1.6
for the clustered inputs with $M=3$ and $T_{\rm i}=20$ msec. 
Filled and open circles denote the results of the 
transient ($t < 1000$ msec) and 
asymptotic solutions ($t > 1000$ msec), respectively.
The dashed lines  are expressed by the equations 
written beside the lines.
The enlarged plots of the regions between dotted, 
vertical lines in Fig. 3(b) are shown 
in Figs. 4(a)-4(c) (see text).

\vspace{0.5cm}

\noindent
{\bf Fig.4}
The enlarged plot of the $\tau_{\rm d}$ dependence 
for (a) $10 \siml \tau_{\rm d} \siml 20 $ msec, 
(b) $20 \siml \tau_{\rm d} \siml 30 $ msec, and
(c) $16 \siml \tau_{\rm d} \siml 20 $ msec
for $M=3$, $T_{\rm i}=20$ and $c=1.6$@
(see Fig. 3(b)).

\vspace{0.5cm}

\noindent
{\bf Fig.5}
The $c$ dependence of the
distribution of $T_{{\rm o}}$  
for (a) $\tau_{\rm d}=10$ and (b) 13.75 msec 
to the clustered inputs with $M=3$ and $T_{\rm i}=20$ msec.
Filled and open circles denote the results of the 
transient ($t < 1000$ msec) and 
asymptotic solutions ($t > 1000$ msec), respectively.
The enlarged plot of the regions between dotted, 
vertical lines in Fig. 5(b) is shown in Fig. 6 
\vspace{0.5cm}

\noindent
{\bf Fig.6}
The enlarged plot of the $c$ dependence of the
$T_{\rm o}$ for $M=3$, $T_{\rm i}=20$ 
and $\tau_{\rm d}=13.75$ msec (see Fig. 5(b)). 
\vspace{0.5cm}

\vspace{0.5cm}

\noindent
{\bf Fig.7} 
The time course of sequential input ($U_{\rm i}$), output ($U_{{\rm o}j}$),
the total postsynaptic current ($I_{j}$), and the
membrane potential ($V_{j}$) 
with $T_{\rm i}=20$, $\tau_{\rm d}=10$ msec and $c=1.0$.
The result of $V_2$ is shifted
downward by 200 mV and scales for $U_{\rm i}$, 
$U_{{\rm o}j}$ and $I_j$ are arbitrary.

\vspace{0.5cm}

\noindent
{\bf Fig.8} 
The $\tau_{\rm d}$ dependence of the
distribution of $T_{{\rm o}}$ of (a) $c=1.0$ 
and (b) 1.6 for the sequential input
with $T_{\rm i}=20$ msec. 
%Dashed lines in (b) are given by eq. (17) with a pair of
%integers $(\ell, m)$ shown beside the line.
The enlarged plot of the region between dotted, 
vertical lines in Fig. 8(a) is shown in Fig. 9. 

\vspace{0.5cm}

\noindent
{\bf Fig.9}
The enlarged plot of the $\tau_{\rm d}$ dependence of $T_{{\rm o}}$
at (a) $21 \leq \tau_{\rm d} \leq 26$ msec and 
(b)  $23 \leq \tau_{\rm d} \leq 24$ msec
[the region sandwiched by vertical dotted lines in 9(a)]
for $T_{\rm i}=20$ msec and $c=1.0$ (see Fig. 8(a)).

\vspace{0.5cm}

\noindent
{\bf Fig.10}
The $c$ dependence of the distribution of $T_{{\rm o}}$
for (a) $\tau_{\rm d}$=10.0 and (b) 13.75 msec
to the sequential inputs with $T_{\rm i}$=20 msec. 
The enlarged plot of the regions between 
dotted, vertical lines in 10(b) is shown
in Fig. 11.
\vspace{0.5cm}

\noindent
{\bf Fig.11} 
The enlarged plots of the $c$ dependence of $T_{{\rm o}}$
for $T_{\rm i}=20$ and 
$\tau_{\rm d}=13.75$ msec (see Fig. 10(b)).
The arrow denotes the $c$ value for which the Lorentz plot
is shown in Figs. 12(a).

\vspace{0.5cm}

\noindent
{\bf Fig.12}
(a) The Lorenz plot of $T_{{\rm o}m}$ for $c=0.95$
with $T_{\rm i}=20$ and $\tau_{\rm d}$=13.75, 
the computation being performed for 20 sec (two million steps).
(b) The correlation integral $C(\epsilon)$ of the cylce
shown in (a) as a function of 
$\epsilon$ in the log-log plot for various dimensions $k$,
the dashed line denoting $C(\epsilon) \propto \epsilon^{\nu}$ 
with the correlation dimension of $\nu=0.94$  [eqs.(3.3)-(3.5)].

\vspace{0.5cm}

\noindent
{\bf Fig.13}
The $T_{\rm i}$ dependence of the
distribution of $T_{\rm o}$ for (a) the clustered input ($M=3$)
and (b) sequential spike-train input with $\tau_{\rm d}=50$ msec
and $c=1.0$.
Filled and open circles in (a) denote the results of the
transient ($t \leq 1000$ msec) and 
asymptotic solutions ($t > 1000$ msec), respectively,
while in (b) filled circles express the results of
asymptotic solutions ($t > 1000$ msec).
Dashed lines are expressed by a pair of integers of
($\ell, m$) in eq. (17) (see text).
\vspace{0.5cm}

\noindent
{\bf Fig.14}
The time dependence of $T_{{\rm o}}$ for the clustered impulse
inputs with $M=3$, 10, 50 and $\infty$ with $T_{\rm i}=20$, 
$\tau_{\rm d}=50$ msec and $c=1.0$,
results of $M$=3, 10 and 50 being shifted upward 
by 60, 40 and 20 msec, respectively.
The arrows denote the time
below which the inputs are continuously applied.

\vspace{0.5cm}

\noindent
{\bf Fig.15}
The time dependence of $T_{{\rm o}}$ for the clustered impulse
inputs with $M=3$, 10, 50 and $\infty$ with $T_{\rm i}=20$, 
$\tau_{\rm d}=13.75$ msec and $c=0.95$,
results of $M$=3, 10 and 50 being shifted upward by 30, 20 and 10 msec, 
respectively.

\vspace{0.5cm}

\noindent
{\bf Fig.16}
The Lorenz plots of $T_{{\rm o}m}$ of (a) the single HH neuron
($c=0.0$) and (b) the coupled HH neurons 
($\tau_{\rm d}=10$ msec and $c=1.0$)
receiving spike-train inputs whose ISIs are  modulated
by sinusoidal signal given by eq. (21)
with $d_0=2 d_1=20$ and $T_p=100$ msec
 (see text).

\vspace{0.5cm}

\noindent
{\bf Fig.17}
The time correlation $\Gamma_{12}(\tau)$ between the membrane 
potentials of the neurons 1 and 2 [eq. (22)] for
(a) the constant-ISI input with $T_{\rm i}=20$, 
$\tau_{\rm d}=10$ msec and $c=1.0$,
(b) that with $T_{\rm i}=20$. $\tau_{\rm d}=13.75$ msec 
and $c=0.95$,
and (c) the sinusoidal spike-train input given by eq. (21) 
with $d_1=2 d_2=20$, $T_{\rm p}=100$,
$\tau_{\rm d}=10$ msec and $c=1.0$.
The results of (b) and (c) are shifted downward
by 1.0 and 2.0, respectively (see text).

\vspace{0.5cm}

\noindent
{\bf Fig.18}
The $\tau_{\rm d}$ dependence of the mean ($\mu_{\rm o}$) and
rms ($\sigma_{\rm o}$) of output ISIs of the coupled HH neurons
($c=1$) receiving sequential inputs of 
$\mu_{\rm i}=<T_{\rm i}>=20$ msec
with $c_{\rm vi}=0.0$ (dashed curves) and 
$c_{\rm vi}=0.43$ (solid curves) (see text).  
\vspace{0.5cm}

%------------------------------------

\begin{thebibliography}{99}
\bibitem{note2}
The intial conditions given by eq. (14) and the intitial function
given by eq. (15) are set every time at the beginning of
the calculation when the values of a given set of
parameters 
of $c$, $\tau_{\rm d}$ and $T_{\rm i}$ are changed.
\end{thebibliography}
\end{document}